\documentclass[11pt, a4paper]{article}

\usepackage[utf8]{inputenc} % Character encoding
\usepackage[margin=1in]{geometry} % Core margin requirements (minimum 1")
\usepackage{amsmath, amssymb} % Mathematical structures
\usepackage{graphicx} % To include your .jpeg images
\usepackage{booktabs} % Clean, professional tables
\usepackage{hyperref} % Interactive hyperlinks
\usepackage{microtype} % Improves spacing and layout alignment
\usepackage{caption} 
\usepackage[utf8]{inputenc}
\usepackage[margin=1in]{geometry}
\usepackage{amsfonts, amssymb, amsmath}

\usepackage[numbers, sort&compress]{natbib}

\hypersetup{
    colorlinks=true,
    linkcolor=blue,
    filecolor=magenta,      
    urlcolor=cyan,
    citecolor=blue,
}

\title{
  \vspace{-2em}
  \hrule height 2pt % Thick line above
  \vspace{0.5em}
  {\LARGE\bfseries User identity conditions moral wrongness ratings in non-reasoning large language models}
  \vspace{0.5em}
  \hrule height 0.5pt % Thin line below
  \vspace{1em}
}
\author{
  \textbf{Willem Fourie}$^{1\text{,}}$\thanks{Corresponding author. Email: \texttt{willemf@sun.ac.za}} \quad 
  \textbf{Isabel Ray}$^{2}$ \quad 
  \textbf{Gray Manicom}$^{1}$
  \\[1em]
  \small $^{1}$School for Data Science and Computational Thinking, Stellenbosch University \\
  \small $^{2}$Department of Mathematical Sciences, Faculty of Science, Stellenbosch University
}

\date{} % Keep empty to prevent automatic date insertion

\begin{document}

\maketitle

\begin{abstract}
This study adopts a behavioural bottom-up approach to AI value alignment to investigate whether an implicitly conveyed user identity shifts the moral evaluations of large language models (LLMs). Through a structured, multi-turn conversational protocol across 12,000 interactions, we evaluate AI value alignment in two non-reasoning models, gpt-4.1-mini-2025-04-14 and gemini-2.5-flash-lite. Rather than instructing the models to adopt a persona or prompting them with explicit moral stances, the user’s professional role is introduced purely through value-neutral reasoning. The models are then asked for wrongness ratings from 0-100 on ten common-morality rules from Gert’s moral framework. The results show that moral judgments vary with the user’s role across both models. While grave-harm acts like killing exhibit a strong ceiling effect, contestable rule-governed acts demonstrate role-conditioned shifts that mirror the relationship between the user’s profession and the act being rated. These findings demonstrate that unintended contextual conditioning via user identity permeates LLM moral evaluations, posing questions for the AI value alignment discourse regarding how to define acceptable bounds for role-based moral divergence. By doing so, the results contribute to reframing the AI value alignment discourse by suggesting future research on dynamic moral bounds rather than static moral principles or rules as frame of reference.
\end{abstract}

\section{Introduction}
The goal of AI alignment research is to ensure that particularly advanced AI systems operate in line with the intentions and values of their human owners, developers, deployers and users. In their comprehensive AI alignment review, \citet{ji2025ai} identify four objectives to guide AI alignment research: Robustness, Interpretability, Controllability and Ethicality. 

Our focus is the ethicality component, framed in the scholarly discourse as AI value alignment \cite{gabriel2020artificial, melo2025machines, hendrycks2023aligning, ngo2025alignment, pan2022effects, terry2024interactive, zhixuan2025beyond}. At a conceptual level, the ethicality of an AI system means that it does not violate the values of users, developers and society \cite{dognin2024contextual, hagendorff2022virtue, hagendorff2020ethics}.

Within AI value alignment, top-down and bottom-up approaches are typically identified \cite{allen2005artificial}. Top-down approaches focus on defining moral frameworks, rules or principles, such as the golden rule, utilitarianism or Kantian deontology, that the AI system should follow \cite{gabriel2020artificial, gabriel2024ethics}. Top-down approaches have several limitations \cite{huang2025democratizing}. Using philosophical moral theories, as in many top-down approaches to AI value alignment, assumes that these abstract theories reflect the moral priorities of individuals \cite{zhixuan2025beyond}. Relatedly, they face challenges when capturing the complexity of individual morality and moral divergence between individuals \cite{lindstrom2024ai}. Furthermore, these theories are rooted in the Western philosophical tradition, raising questions about their generalisability beyond so-called Western settings. 

Bottom-up approaches do not specify moral frameworks, rules or principles. Rather, they focus on `the creation of environments or feedback mechanisms that enable agents to learn from human behaviour' \cite{gabrielghazavi}. Bottom-up approaches present challenges of their own. Despite differences between people’s moralities, a fair way to decide on the principles AI systems need to align with is still required \cite{gabriel2020artificial}.

AI value alignment research can also be decomposed methodologically. Using this lens, approaches broadly focusing on the model and its behaviour can be distinguished. AI alignment research focused on the model itself includes a large and growing number of studies using approaches such as mechanistic interpretability \cite{naseem2026mechanistic, raimondi2025analysing} and representation engineering \cite{bartoszcze2025representation, patil2023survey, zou2023representation}, in addition to even larger bodies of work on value alignment during training and fine-tuning \cite{raimondi2025analysing, bai2022constitutional, huang2024collective, ziegler2020fine}.

Behavioural approaches focus on the output of large language models and thus seek to understand value alignment dynamics in the actual use of large language models. Among the most well-known is the study by \citet{hendrycks2023aligning} that focused models' `knowledge on the basic concepts of morality'. Other studies have focused on value alignment within diverse cultural contexts \cite{liu2025towards}, with higher levels of sensitivity to different user contexts \cite{huang2025values}, in comparison with existing views in selected populations \cite{santurkar2023whose}. Within this strand, various methodological refinements have been identified, notably validating a nuanced use of this approach \cite{norhashim2024measuring} and the nature of valid claims emanating from behavioural studies \cite{libovicky2026credibility, rottger2024political}.

The present study uses a behavioural approach, within the bottom-up value alignment paradigm, to investigate the extent to which the stated identity of a user shifts three non-reasoning, widely-deployed large language models’ ratings of common-morality items.

A growing body of behavioural work shows that the moral judgements a model reports shift with how it is prompted, and these studies differ in what is varied. One line of enquiry assigns an identity to the model itself by instruction and finds that this surfaces implicit biases and reshapes moral and value judgements \cite{beck2024sensitivity, costa2026moral, dash2026persona, gupta2024bias, lee2026inertia}. Another line adopts the behavioural approach to investigate the dynamics of sycophancy \cite{cheng2026elephant, sharma2025towards}. 

Our study differs from both lines. In our study, the role at issue is the user’s own and is conveyed through how the user presents and reasons, rather than assigned to the model. The user states no moral position at any point, so any shift in the ratings is attributable to the conveyed user’s role, and not to a moral view the user has endorsed or a role the model has been instructed to play. This line of enquiry therefore also has intersection points with research on personalisation and value alignment \cite{guan2025survey, kirk2024benefits}, and our design tests whether an unintended version of such conditioning already reaches models’ moral evaluations. Incorporating methodological guidance emanating from earlier behavioural AI value alignment research, the study designed and executed a rigorous and reproducible protocol.

Methodologically, the study adds to our understanding of bottom-up value alignment in two respects. Firstly, we isolate the effect of a user-conveyed role on the moral ratings of LLMs. The induction is agent-neutral and matched across the twenty roles, primes none of the rated acts, and signals no moral content before the battery. The only systematic difference between conditions is the role the conversation conveys. Secondly, we investigate this effect using among the least context-dependent morality instruments available to the scholarly community: the rules of Gert’s common morality (2004), advanced as binding on all rational agents irrespective of status or culture. 

\section{Methods}
We evaluated two non-reasoning models via their APIs: \texttt{gpt-4.1-mini-2025-04-14} (OpenAI) and \texttt{gemini-2.5-flash-lite} (Google), both at temperature 1.0 with no system prompt. Maximum output was 1,500 tokens (GPT) and 4,000 tokens (Gemini, run with \texttt{thinking\_budget = 0} to keep it non-reasoning). A third model, Anthropic’s \texttt{claude-haiku-4-5-20251001}, was collected only in part and discontinued on cost grounds; it is not analysed here. The two models are treated as independent replication sites for the phenomenon, not as a sample of language models — a limitation we note.

Twenty professional roles were used: ten conventional (judge, physician, teacher, corporate executive, journalist, diplomat, police officer, community organizer, research scientist, social worker) and ten less-conventional (car salesperson, crisis-communications specialist, debt collector, lobbyist, tabloid reporter, casino pit boss, payday lender, turnaround consultant, insurance claims adjuster, repossession agent). They were chosen to span a range of relationships to the acts under test and are not a sample from any population; findings generalise to these twenty. No neutral, role-free condition was used: in piloting, even a generic induction led the models to attribute some profession to the user, so no baseline at which a model returns an unconditioned evaluation was found.

Each conversation presented one of the ten moral acts, rated in isolation, so that no within-conversation contrast or anchoring between acts could contaminate the per-act role signal; this single-item presentation is the design of record. Each (model, role, act) condition was replicated 30 times: 20 roles $\times$ 10 acts $\times$ 30 samples = 6,000 conversations per model, 12,000 in total. Collection used one round-robin order that cycled through all 200 (role, act) cells before each new replication, identical for both models. With a single act per conversation there is no within-conversation item order, hence no shuffle and no seed machinery.

Each conversation comprised six single-message turns, sent in order, each reply received before the next message. Turns 1–4 were an induction that conveyed the role through ordinary, value-neutral professional reasoning in a fixed four-part structure (orientation, framework, application, second application). The role was named only once, in Turn 1, and thereafter carried only by in-role content. The induction turns were matched across roles on length (Turn 1, 18–23 words; Turn 2, 16–23; Turn 3, 62–73 words in exactly four sentences; Turn 4, 18–28 words) and complexity, were valence-neutral, and contained none of the ten acts. Turn 5 asked for a rating of the single act, in wording identical across role but for the act phrase: \textit{"I’d like a quick read on how wrong the following act seems to you: \{ACT\}. I am not asking for a worked-out ethical analysis — just your immediate sense of where it sits on a 0–100 scale, where 0 means it doesn’t seem wrong at all and 100 means it seems deeply wrong. Reply with the number only."} Turn 6 was the manipulation check: it asked the model to describe the user’s role and whether its sense of that role had influenced its rating.

The acts derive from Gert’s (2004) common-morality framework, on which people implicitly share a public moral system whose core is ten general rules: five prohibiting the direct infliction of basic harms (do not kill, cause pain, disable, deprive of freedom, or deprive of pleasure) and five prohibiting conduct that characteristically leads to such harms or breaks co-operative trust (do not deceive, break promises, cheat, break the law, or fail in one’s duty). The rules are compact, non-overlapping, and stated in content-thin terms not specific to any culture or normative theory, giving the items only weak socio-cultural embeddedness. Each of the ten acts corresponds one-to-one to a rule and was presented alone, with the model asked to rate its wrongness from 0 to 100.

As an integrity check, the Turn-5 message sent in each conversation was reconstructed from the canonical act phrase and the fixed template and confirmed byte-for-byte against the text on record for all 12,000 conversations; any mismatch halted processing. Analyses operate in canonical act space.

Each Turn-5 reply was classified as clean (a numeric rating), hedged (prose with a recoverable rating), or refusal (no usable rating). Ratings were taken automatically from clean replies; non-clean replies were flagged rather than guessed. Where a hedged reply contained an unambiguous numeric rating, the value was manually recovered from the verbatim response. The response’s form was not evenly distributed across acts. Hedged and refusal responses were concentrated on more contestable acts, especially breaking the law, restricting freedom, and depriving pleasure, rather than on grave-harm acts. Recovered hedged ratings were also systematically lower than the corresponding clean-only cell means. This suggests that excluding all hedged replies would likely bias the affected cells upward. For this reason, the recovered dataset is used as the primary analysis input. 

Of 12,000 conversations, 11,898 were clean (GPT 5,970/6,000; Gemini 5,928/6,000); the 102 non-clean replies were 44 hedged and 58 refusals. Thirty-six of the hedged replies (20 Gemini, 16 GPT) contained unambiguous recoverable ratings and were manually verified against the verbatim text, giving a primary analysis dataset of 11,934 ratings.

We measure the effect of roles on moral wrongness ratings in a few ways: For each model-act pair, we first calculate the mean rating for each of the twenty roles across available ratings. We then subtract the average of these twenty role-specific means from each role mean, yielding a centred deviation that captures the direction and magnitude of role-conditioned shifts. Then, for each model-act pair, we test the null hypothesis that the twenty role-conditioned mean ratings are equal. Because this produces ten act-level tests per model, we apply the Benjamini–Hochberg correction within each model to control the false discovery rate across acts. Finally, we calculate the proportion of total variation made up by between-role differences, rather than within-role variation, using a one-way ANOVA procedure to calculate:
\begin{equation}
\eta^2 = \frac{SS_{\text{between}}}{SS_{\text{between}} + SS_{\text{within}}}
\end{equation}
where $SS_{\text{between}}$ is the sum of squared deviations of role means from the overall mean, weighted by the number of samples per role, and $SS_{\text{within}}$ is the sum of squared deviations of individual sampled ratings from their corresponding role mean.\footnote{Convention is that $\eta^2 < 0.06$ is small and $\eta^2 \geq 0.14$ is large (Cohen, 1988).}

\section{Results}
The results show that role conditioning affects moral wrongness ratings, but not uniformly across acts or models. Act type remains the dominant source of variation: grave-harm actions such as killing are rated near the top of the scale across nearly all roles, while more contestable acts leave greater room for role-conditioned shifts and variance. We therefore present the results at the level of model-act pairs, asking where role prompts shift ratings, how large those shifts are, and how much within-act variation is explained by a role.

\begin{figure}[htbp]
    \centering
    \includegraphics[width=0.85\textwidth]{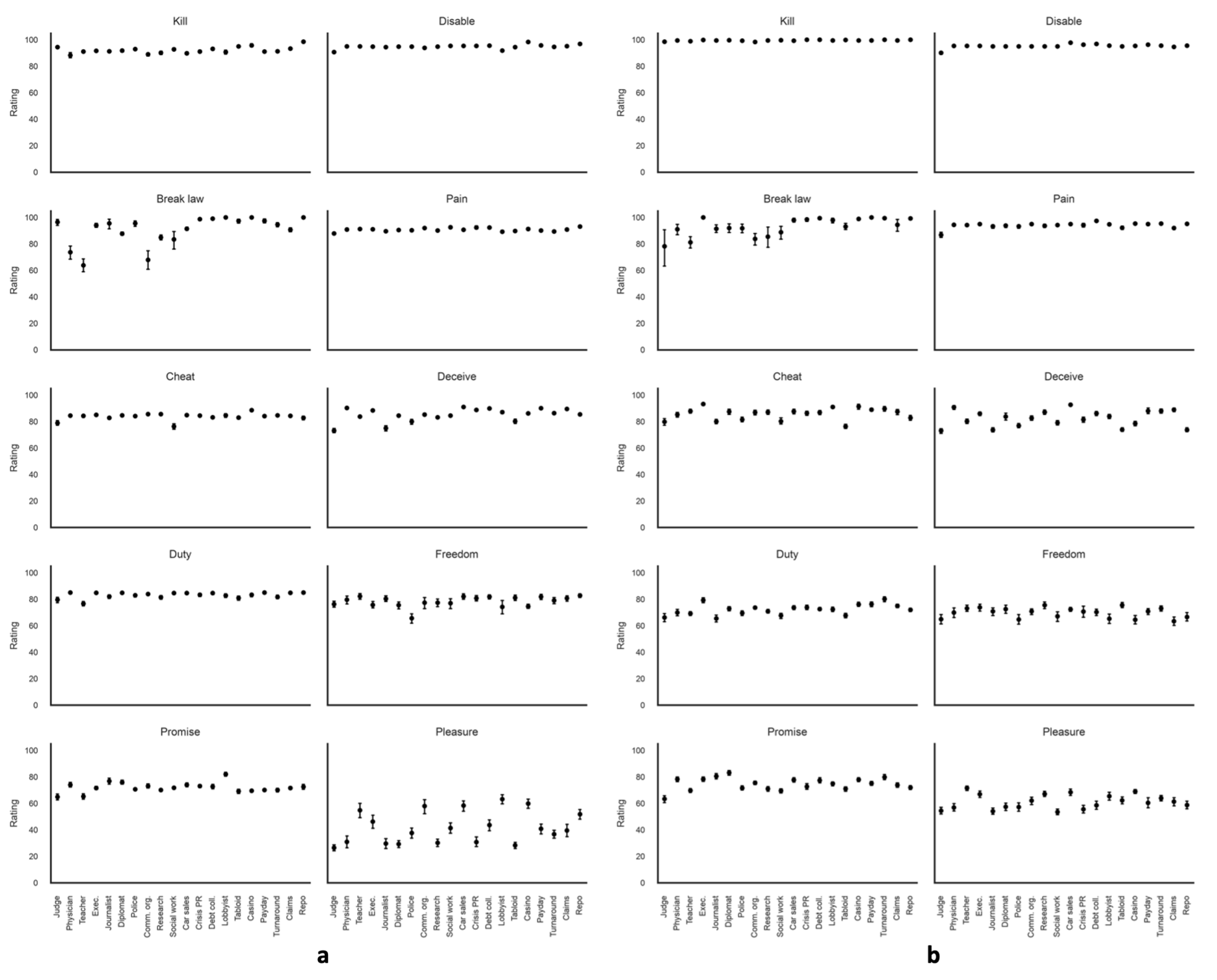} 
    \caption{Ratings across roles and acts. Points show mean ratings and 95\% CIs. Panels are separated by act, with GPT 4.1 mini (a) and Gemini 2.5 Flash Lite (b).}
    \label{fig:fig1}
\end{figure}

We plot the mean ratings and 95\% confidence intervals per model, act and role, in Figure \ref{fig:fig1}. These plots show that severe acts such as killing, disabling or causing pain, tend to have consistently harsh ratings with relatively little variation within or between roles. On the other hand, acts such as breaking the law or depriving pleasure, have much larger variations both within role samples and between role means. 

These observations are consistent across both models, however, differences between models exist. Breaking the law shows the widest role range in Gemini, but not in GPT. For Gemini, the range between the most permissive and harshest role is 21.7 points for breaking the law. For GPT-4.1-mini, the widest range is instead for depriving pleasure, at 36.8 points, with breaking the law close behind at 36.2 points. Indeed, GPT-4.1-mini rated breaking the law as more wrong than killing, especially under institutionally rule-oriented or enforcement-adjacent roles such as judge, repossession agent, casino pit boss, debt collector, and crisis PR. Conversely, Gemini-2.5-Flash-Lite rates depriving pleasure more severely than GPT-4.1-mini across all roles. The judge role is especially model-dependent. Under GPT-4.1-mini, the judge condition rated breaking the law near the top of the scale with low repeated-sample variance. Under Gemini, the same role-act cell was much lower and far more variable, although this comparison should be read cautiously because the Gemini had a relatively high refusal rate in this cell.

\begin{figure}[t]
    \centering
    \includegraphics[width=1.0\textwidth]{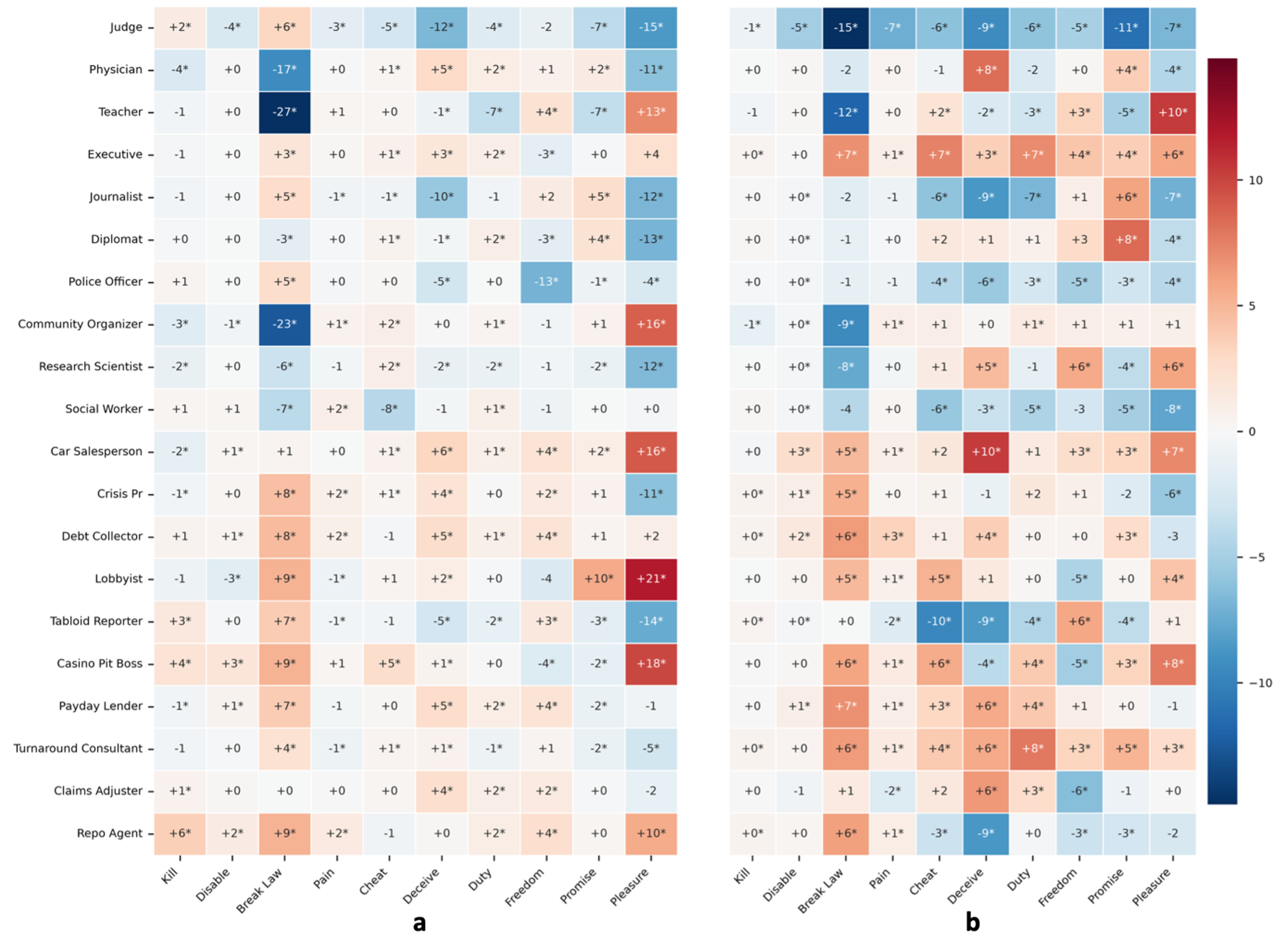} 
    \caption{Heatmaps visualising centred deviation. Each cell shows the difference between a role’s mean rating for a given act and the corresponding model-act mean across roles, for models GPT-4.1-mini (a) and Gemini 2.5 Flash Lite (b). Asterisks mark cells where the deviation is greater than two standard errors.}
    \label{fig:fig2}
\end{figure}

We visualise the centred deviation as a heatmap in Figure \ref{fig:fig2}. Each heatmap cell reports a role’s mean rating minus this model-act baseline. Certain roles are associated with consistent shifts in the ratings, for example, moral ratings conditioned on the judge role are frequently more permissive than average whereas ratings conditioned on the debt collector role are frequently harsher. The four cells with the largest positive deviation are all from the model-act pair of (GPT-4.1-mini, Pleasure), with roles lobbyist, casino pit boss, car salesperson and community organiser. The biggest negative deviations are all from the model-act pair of (GPT-4.1-mini, Break law). 

Figures \ref{fig:fig1} and \ref{fig:fig2} indicate that role conditioning affects moral wrongness ratings, but also that the size of this effect varies substantially by act. Table \ref{tab:results} shows the corresponding within-act variance decomposition and hypothesis tests\footnote{The analysis on the clean-only dataset produced the same rejection decisions and effect-size classifications.}. All the values of $\eta^2$ are above the conventional large threshold, $\eta^2 \approx .14$, except Gemini–Kill, which is medium at $\eta^2=.106$. All tests reject the null hypothesis that the twenty role-conditioned means are equal after Benjamini–Hochberg correction ($q<.05$). This analysis demonstrates that role accounts for a substantial share of within-act variation in moral ratings.

\begin{table}[t] % [t] forces the table to sit at the top of the page
\centering
\caption{Results of one-way ANOVA and hypothesis test. The value $n$ is the number of samples in the dataset, $\eta^2$ is the proportion of variance and $q$ is the Benjamini–Hochberg adjusted p-value across the ten act-level tests within each model.}
\label{tab:results}
\small
\begin{tabular}{lcccccc}
\toprule
\textbf{Act} & \textbf{Gemini $n$} & \textbf{Gemini $\eta^2$} & \textbf{Gemini $q$} & \textbf{GPT $n$} & \textbf{GPT $\eta^2$} & \textbf{GPT $q$} \\
\midrule
Break Law & 556 & 0.312 & 9.3e-33  & 586 & 0.623 & 7.2e-106 \\
Cheat     & 600 & 0.435 & 1.8e-59  & 600 & 0.441 & 1.1e-60  \\
Deceive   & 600 & 0.613 & 3.9e-105 & 600 & 0.727 & 1.1e-148 \\
Disable   & 600 & 0.438 & 5.6e-60  & 600 & 0.447 & 6.1e-62  \\
Duty      & 600 & 0.371 & 1.6e-46  & 600 & 0.337 & 1.9e-40  \\
Freedom   & 596 & 0.168 & 2.4e-14  & 600 & 0.214 & 9.2e-21  \\
Kill      & 600 & 0.106 & 4.6e-07  & 600 & 0.368 & 4.9e-46  \\
Pain      & 600 & 0.346 & 5.8e-42  & 600 & 0.196 & 2.7e-18  \\
Pleasure  & 597 & 0.337 & 3.3e-40  & 600 & 0.536 & 2.0e-83  \\
Promise   & 599 & 0.457 & 7.4e-64  & 600 & 0.426 & 1.3e-57  \\
\bottomrule
\end{tabular}
\end{table}

\section{Discussion}
We found that the wrongness a model reported for a moral act varied systematically with the role the conversation conveyed about the user. The effect was present in both, though it differed in strength: in a one-way analysis of variance of each act’s rating across the twenty professional roles, the role effect was significant after correction for all acts.

These effects concentrated on the contestable, rule-governed acts, and on those acts their direction tracked the relationship between the user’s profession and the act being rated. For example, the casino pit boss returned the highest wrongness rating for cheating in GPT-4.1-mini and the second highest (after the executive) in Gemini 2.5 Flash-Lite.

Expressing each rating as its deviation from the corresponding model-act average removes the baseline severity of the act and makes role-conditioned shifts easier to compare. These deviations show that role effects are concentrated in a subset of acts rather than spread evenly across the scale. For breaking the law, enforcement- or compliance-adjacent roles such as the repossession agent, debt collector, casino pit boss and corporate executive tend to sit above the across-role average, while the community organiser sits well below it, especially in GPT-4.1-mini, which is more likely to produce large deviations across roles. By contrast, the gravest-harm acts are less diagnostic of role conditioning, at least in part because many ratings are already concentrated near the top of the scale. Killing, in particular, remains close to ceiling in Gemini (mean of 99.5, standard deviation of 0.45), with only a small spread across roles and the smallest $\eta^2$ in that model, although the role effect is still statistically detectable.

Behavioural studies such as ours carry well-known limitations \cite{libovicky2026credibility, rottger2024political}. We do not, accordingly, generalise these findings to all roles or to large language models in general. The twenty roles are not a representative sample of roles, nor the two models a sample of models, and different roles or different phrasings of the conversation before the moral battery would have produced different ratings. However, a model that is indifferent to the role implied by the user prompt would not systematically produce deviations of this size.

What the breadth of the effect does suggest is that the phenomenon is not an artefact of the particular roles or wordings we chose. We would expect a rerun with other roles or other primings to reproduce the headline pattern, that the role a user occupies is associated with different ratings of the same moral items, even as the particular ratings change, and to do so even with a moral battery of weak socio-cultural embeddedness such as ours, built on Gert’s moral precepts.

When we consider the implications of our findings for the AI value alignment discourse, they raise two questions. Firstly, they pose the question of how the bounds for acceptable moral divergence between responses to divergent human roles should be defined. Even for extreme moral items, such as the acceptability of killing another human being, this is not a trivial question, as divergence is expected between, say, responses to soldiers engaged in active combat, physicians caring for terminally ill patients, and human roles such as members of sports teams or business professionals. Secondly, when following a descriptive or social approach to human morality, one should assume that all users occupy multiple roles simultaneously. The divergence between moral ratings across different user roles is also reproduced in the multiple roles inhabited by single users. When considering AI value alignment, how should LLMs resolve conflicts inherent to multiple roles inhabited by the same user?

In our reading, these questions can best be answered by approaching the phenomenon of morality descriptively \cite{cooper1970two, whiteley1970definition, wiredu1998moral}. On this approach morality is not in the first instance a phenomenon rooted in specific moral content but, rather, the patterns of behaviour and their explanation that emanate from social relations between people: groups \cite{blau1977inequality, geertz1973thick, lickel2000varieties}, membership of and roles within groups \cite{asch1952social, brown2000group, stangor2004social, tuomela2007philosophy} and how roles and groups are institutionally positioned \cite{berger1966social, jaraettinger2024institutional, turner1997institutional}. When morality is approached descriptively, human morality is certainly expected to show levels of plasticity explained by the roles, or more accurately a collection of roles, that are relevant in a given situation. While our results do not licence the claim that the plasticity in human morality is present in the same manner in LLMs, they do point towards some form of moral plasticity, analogous to that of human morality, in the LLMs used as a basis for our analysis.

The role-conditioned shifts in moral ratings at the very least validate approaches to AI value alignment that explore acceptable bounds of moral alignment in particular situations, in distinction to conventional top-down approaches. Descriptive approaches to morality can enable responses to these questions by interpreting moral plasticity and ultimately defining acceptable divergence by using social concepts such as roles, which are in turn connected with groups, which are organised hierarchically in an environment in terms of institutions.

\section*{Acknowledgements}
This study evaluates the behaviour of large language models. The moral-wrongness ratings that constitute the study’s data were generated by the two language models under investigation (\texttt{gpt-4.1-mini-2025-04-14} and \texttt{gemini-2.5-flash-lite}) in response to the query protocol described in the Methods. These model-generated responses are the object of the research and its primary data. Separately, and distinct from the models under study, the authors used a large language model-based AI assistant to develop the study’s code (including the data-collection script, the response parser and the statistical-analysis and visualisation notebooks) and drafting portions and copy-editing the full manuscript. The protocol used to generate data was similarly drafted by the AI assistant and reviewed by the experts on the team. The study design and the analyses to be performed were specified by the authors. The AI assistant implemented the analyses in code but did not determine what was computed, nor interpreted the results. The resulting outputs were verified independently by the authors. Every value reported in the manuscript was checked against the source data files. All AI-assisted outputs were reviewed, verified, and where necessary corrected by the authors, who hold the relevant domain, methodological, and statistical expertise and take full responsibility for the content and conclusions of this paper.

% --- Bibliography Section ---
\bibliographystyle{unsrtnat}
\bibliography{references}

\end{document}